\documentclass[letterpaper, 10 pt, conference]{ieeeconf}  
\usepackage[left=54pt,right=54pt,top=54pt,bottom=54pt]{geometry}
\IEEEoverridecommandlockouts                              
\usepackage{graphics} 
\usepackage{epsfig} 
\usepackage{mathptmx} 
\usepackage{times} 
\usepackage{amsmath} 
\usepackage{amssymb}  
\usepackage{booktabs} 
\usepackage{cite}
\usepackage{amsfonts}
\usepackage{amsbsy}
\usepackage{placeins}
\usepackage{multirow}
\usepackage{tabularx}
\usepackage{bm}
\usepackage{url}

\title{\LARGE \bf
Improving Initial Transients of Online Learning Echo State Network Control System with Feedback Adjustments}

\author{Junyi Shen
\thanks{Junyi Shen is with The University of Tokyo, 7-3-1 Hongo, Bunkyo-Ku, Tokyo, Japan
        {\tt\small junyi-shen@g.ecc.u-tokyo.ac.jp}}
}

\begin{document}

\maketitle
\thispagestyle{empty}
\pagestyle{empty}

\begin{abstract}
Echo state networks (ESNs) have become increasingly popular in online learning control systems due to their ease of training. However, online learning ESN controllers often suffer from slow convergence during the initial transient phase. Existing solutions, such as prior training, control mode switching, and incorporating plant dynamic approximations, have notable drawbacks, including undermining the system's online learning property or relying on prior knowledge of the controlled system. This work proposes a simple yet effective approach to address the slow initial convergence of online learning ESN control systems by integrating a feedback proportional-derivative (P-D) controller. Simulation results demonstrate that the proposed control system achieves rapid convergence during the initial transient phase and shows strong robustness against changes in the controlled system's dynamics and variations in the online learning model's hyperparameters. We show that the feedback controller accelerates convergence by guiding the online learning ESN to operate within a data range well-suited for learning. This study offers practical benefits for engineers aiming to implement online learning ESN control systems with fast convergence and easy deployment.
\end{abstract}

\section{INTRODUCTION}

Academia and industry have long collaborated on developing nonlinear control algorithms that are both effective and easy to implement. While model-based control approaches are widely preferred due to their analyzability and capability to control nonlinear systems \cite{cheng2016adaptive, zou2023generalized}, accurately modeling complex dynamic systems continues to pose a significant challenge \cite{waegeman2012feedback}. On the other hand, proportional-integral-derivative (PID) controllers, which are extensively employed in model-free control systems due to their simplicity of implementation, often fall short of delivering satisfactory performance when applied to nonlinear systems \cite{liang2023online}. Recurrent neural networks (RNNs), capable of capturing complex nonlinear dynamics such as the hysteresis properties in soft actuators \cite{sun2022physics}, have been employed as the feedforward element in controlling nonlinear plants \cite{jiang2022intelligent}. However, the training process for RNNs can be computationally demanding, with large amounts of data and substantial time required.

Reservoir computing models (RCs), such as the widely known echo state network (ESN) \cite{jaeger2002adaptive}, represent a variant of recurrent neural networks (RNNs). In RCs, inputs are projected into a high-dimensional state space with an input layer and a large network known as the reservoir, which typically comprises tens to thousands of nodes \cite{park2016online}. The high-dimensional state is combined with the linear weights in an output layer to generate the model's output. RCs offer a simpler training process than traditional RNNs: the reservoir and the input layer are randomly initialized, scaled by designed strategies, and remain fixed during training. Only weights in the output layer are adjusted by linear regression methods. This gradient-free approach makes RCs favored in online learning control systems \cite{xing2012modeling, liang2023online}. Previous studies have demonstrated the effectiveness of using ESNs with online learning algorithms such as the recursive least squares (RLS) or filtered-X to control nonlinear systems, including hydraulic excavators \cite{park2016online}, oil production platforms \cite{jordanou2019online}, electro-hydraulic shaking tables \cite{liang2023online}, and other challenging tasks \cite{waegeman2012feedback}.

However, online learning ESN controllers often exhibit undesirable initial transients with slow convergence \cite{jordanou2019online}, which can be detrimental in real-world applications \cite{liang2023online}. Therefore, addressing the initial transients of online learning ESN control systems is crucial for their practical implementation. While \cite{waegeman2012feedback} and \cite{jordanou2019online} highlighted the issue of unpleasant initial transients in online learning ESNs, they did not propose specific solutions. In contrast, \cite{park2016online} employed two different methods to mitigate these transients in their simulations and experiments. In their simulations, the ESN was pre-trained before being used for control, thereby eliminating the initial transient of the untrained ESN. In their experiments, a mode-switching mechanism was used where a proportional-derivative (PD) controller initially controlled the system while the RLS algorithm adjusted the ESN's weights. During this phase, the ESN was not involved in controlling the plant. After a predefined period, control was switched to the ESN, which then replaced the PD controller to continue controlling the system. The pre-training in the simulations of \cite{park2016online} makes the system more of an adaptive control system, similar to that in \cite{xing2012modeling}, rather than an authentic online learning system. Similarly, the experimental approach in \cite{park2016online} also compromised the system's online learning nature, as the ESN was adjusted before its application. \cite{liang2023online} adopted a filtered-X method instead of the RLS algorithm to update the online learning ESN and achieved fast convergence at the system's initial stage. However, the filtered-X method in \cite{liang2023online} relies on an approximate nonlinear plant model, which requires prior knowledge of the plant's dynamics or a system identification process. Both can be challenging when the plant exhibits complex and highly nonlinear dynamics. The limitations of existing methods highlight the need for a simpler yet effective strategy to address the online learning ESN control systems' slow initial convergence without compromising their zero-shot online learning property or relying on prior knowledge of the plant's dynamics.

This work shows that the slow convergence issue of online learning ESN control systems can be effectively addressed by simply introducing a feedback controller. Simulations validated the efficacy of this approach and its robustness against changes in the system's dynamics and hyperparameter setups. We show that the feedback controller's role in facilitating the online learning ESN's fast convergence stems from its ability to guide the ESN to operate within a suitable input-output data field for learning. The subsequent content is organized as: Section 2 introduces the design of control system; Section 3 presents the simulation results; Section 4 discusses the feedback controller's contribution; Section 5 concludes.

\section{Control System Design}

\subsection{Online Learning ESN Control System}

The online learning ESN control system typically consists of two identical (except for the input signals) ESN models \cite{waegeman2012feedback}: an ESN-C model for controlling the plant and an ESN-L model for online adjustment. Training neural networks with directly reversed input-output pairs may not yield the desired outcome \cite{zhou2020deep}. Previous works have used future reference signals \cite{waegeman2012feedback, park2016online, jordanou2019online}, which are often available in practical applications \cite{zhou2020deep}, as input to the ESN-C model. In our control system, the ESN-C's input is $\mathbf{y}^\text{d} = [y^d_{k+1}, y^d_{k+2}, \cdots, y^d_{k+\delta}]^\top \in \mathbb{R}^\delta$, which is the future information of the reference signal $y^d$, with a tap size of $\delta$. The input $\mathbf{y}^\text{d}$ is then projected into the high-dimensional state $\mathbf{x}^\text{C} \in \mathbb{R}^r$ in the ESN-C as follows:
\begin{equation}
    \mathbf{x}^\text{C}_{k+1} = (1-\gamma)\mathbf{x}^\text{C}_{k} + \gamma \tanh(\mathbf{W}^\text{r}\mathbf{x}^\text{C}_{k} + \mathbf{W}^\text{in}\mathbf{y}^\text{d}_k),
\label{eq: ESN-C update}
\end{equation}
where $\gamma \in [0,1)$ is known as the leaky rate, $\mathbf{W}^\text{r} \in \mathbb{R}^{r\times r}$ is the reservoir with the size of $r$, $\mathbf{W}^\text{in} \in \mathbb{R}^{r\times \delta}$ is the input layer, and $\tanh(\cdot)$ denotes the hyperbolic tangent activation function that has been commonly used in ESNs. Both $\mathbf{W}^\text{r}$ and $\mathbf{W}^\text{in}$ are randomly initialized, scaled according to respectively specified strategies, and kept fixed throughout model training and execution. The ESN-C's output $u^\text{f}$ is produced by combining the output layer $\mathbf{w}^\text{out} \in \mathbb{R}^{r}$ with the state vector $\mathbf{x}^\text{C}$ as:
\begin{equation}
    u^\text{f}_{k} = \mathbf{w}^\text{out}_{k}\mathbf{x}^\text{C}_{k}.
\label{eq: ESN output}
\end{equation}
Weights in the output layer $\mathbf{w}^\text{out}$ are randomly initialized, adjusted online in the ESN-L model with the RLS algorithm, and subsequently shared with the ESN-C in each step $k$ \cite{waegeman2012feedback}. 

The ESN-L model has identical parameters as the ESN-C model, including $\gamma$, $\mathbf{W}^\text{r}$, $\mathbf{W}^\text{in}$, and $\mathbf{w}^\text{out}$. Differently, the ESN-L model's input is $\tilde{\mathbf{y}}_k = [\tilde{y}_{k-\delta+1}, \tilde{y}_{k-\delta+2}, \cdots, \tilde{y}_{k}]^\top \in \mathbb{R}^\delta$, which is the history of $\tilde{y}$, where $y$ is the plant's actual output and $\tilde{\cdot}$ denotes the signal with feedback sensor noise. In the ESN-L model, the input $\tilde{\mathbf{y}}$ is projected into the high-dimensional state $\mathbf{x}^\text{L} \in \mathbb{R}^r$ as follows:
\begin{equation}
    \mathbf{x}^\text{L}_{k+1} = (1-\gamma)\mathbf{x}^\text{L}_{k} + \gamma \tanh(\mathbf{W}^\text{r}\mathbf{x}^\text{L}_{k} + \mathbf{W}^\text{in}\tilde{\mathbf{y}}_k).
\label{eq: ESN-L update}
\end{equation}
The output layer $\mathbf{w}^\text{out}$ is adjusted by the RLS algorithm as:
\begin{align}
    e^\text{ESN}_{k} &= \mathbf{w}^\text{out}_{k-1}\mathbf{x}^\text{L}_{k} - \bar{u}_{k-\delta} \\
    \mathbf{w}^\text{out}_{k} &= \mathbf{w}^\text{out}_{k-1}-e^\text{ESN}_{k}\left(\mathbf{P}_{k}\mathbf{x}^\text{L}_{k}\right)^\top \\
    \mathbf{P}_{0} &= \frac{\mathbf{I}}{\alpha}  \\
    \mathbf{P}_{k} &= \frac{\mathbf{P}_{k-1}}{\lambda} - \frac{\mathbf{P}_{k-1}\mathbf{x}^\text{L}_{k} \ {\mathbf{x}^\text{L}_{k}}^\top\mathbf{P}_{k-1}}{\lambda\left(\lambda + {\mathbf{x}^\text{L}_{k}}^\top \mathbf{P}_{k-1} \mathbf{x}^\text{L}_{k}\right)}
\end{align}
where $\bar{u}$ denotes the actual input to the controlled system (plant), $0 < \alpha$ is the learning rate, $0 \ll \lambda < 1$ is the forgetting factor, $\mathbf{I} \in \mathbb{R}^{r \times r}$ is the identity matrix, and $\mathbf{P}_{k}$ is the running estimate of the Moore-Penrose pseudoinverse matrix $\left({\mathbf{x}^\text{L}_{k}}^\top \mathbf{x}^\text{L}_{k} + \alpha \mathbf{I} \right)^{-1}$ \cite{waegeman2012feedback}. Smaller values of $\alpha$ enable faster learning but may lead to instability in the RLS algorithm due to rapid weight updates. Similarly, smaller values of $\lambda$ can enhance the algorithm's ability to adapt to changes in the model but may compromise steady-state performance. The output layer $\mathbf{w}^\text{out}$, updated in the ESN-L model, is then transferred to the ESN-C model to perform the operation shown in~(\ref{eq: ESN output}). The stability analysis of the RLS-based online learning ESN control system has been previously discussed in detail by \cite{waegeman2012feedback} and \cite{park2016online} and thus is omitted here for brevity.

\subsection{Combining ESN with Feedback Adjustments}

\begin{figure}[tb]
    \centering
\includegraphics[width=0.85\linewidth]{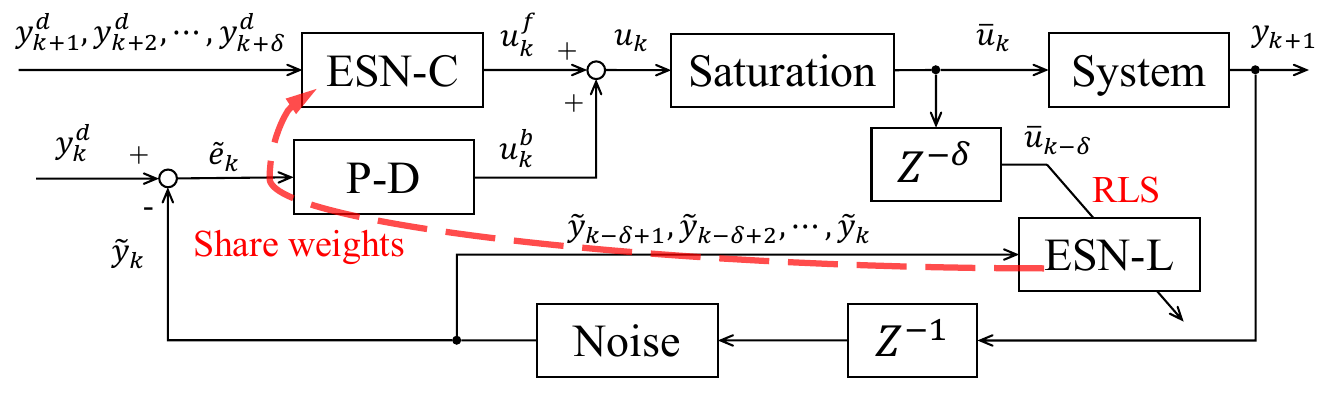}
    \caption{Online learning ESN control system with feedback P-D controller.}
    \label{fig: system}
\end{figure}

Fig.~\ref{fig: system} shows the control system design that integrates the aforementioned online learning ESN control system as the feedforward control element with a P-D controller as the feedback element. The P-D controller processes the control error $\tilde{e}_k = y^d_k - \tilde{y}_k$ to generate the feedback adjustment $u^\text{b}$ as:
\begin{equation}
    u^\text{b}_{k} = K_P \ \tilde{e}_k + K_D \left(\tilde{y}_{k-1} - \tilde{y}_{k}\right),
\end{equation}
where $0 \leq K_P$ is the proportional (P) gain, and $0 \leq K_D$ is the differential (D) gain. The feedback adjustment $u^\text{b}$ is then combined with the ESN-C model's feedforward control output $u^\text{f}$ to form the overall control system output $u$ as:
\begin{equation}
    u_{k} = \bar{u}_{k-1} + u^\text{f}_{k} - u^\text{f}_{k-1} + u^\text{b}_{k},
\label{eq: ESN+FB output}
\end{equation}
the obtained $u_{k}$ is then processed by a saturation function $f_\text{sat}(\cdot)$ to produce the final input to the controlled system as:
\begin{equation}
    \bar{u}_{k} = f_\text{sat}(u_{k}).
\end{equation}

It is important to note that while the system design shown in Fig.~\ref{fig: system} shares similarities with the design in \cite{xing2012modeling}, there are two key distinctions: (1) The ESN controller in \cite{xing2012modeling} is trained offline and adjusted during online execution, making it an adaptive ESN controller, whereas our design is a zero-shot online learning control system. (2) \cite{xing2012modeling} focused on improvement brought by the ESN feedforward to the feedback system's performance, while this work explores how the feedback controller facilitates the fast convergence of the online learning ESN model. As the ESN in \cite{xing2012modeling} was trained before its use, it inherently did not experience the slow initial convergence issue reported in \cite{waegeman2012feedback} and \cite{jordanou2019online}, which the slow convergence is this work's primary emphasis.

\section{Simulation Results}

\subsection{Simulation Setup}

We conducted simulations to validate the control system design shown in Fig.~\ref{fig: system} to control a nonlinear system (plant) output following various reference signals. The simulations were executed on an Apple Silicon M1 CPU using Python 3.10.0 and Numpy 1.23.5. The design of the nonlinear system's dynamics was motivated by the simulation setup in \cite{park2016online}, but the system in this work has greater complexity and nonlinearity, as described by the following equation:
\begin{equation}
    2 {y_{k+1}}^3 = \frac{{y_{k}}^4}{5+{y_{k-1}}^6} + \left( 7 \tanh(\bar{u}_k) \right)^8.
\label{eq: Nonlinear 1}
\end{equation}
The noise in the control system's feedback was simulated as:
\begin{equation}
    \tilde{y}_k = y_k + \mathcal{N}(0, 0.01), 
\end{equation}
where $\mathcal{N}(a, b)$ represents a random value following a normal distribution with the mean of $a$ and the standard deviation of $b$. The saturation function $f_\text{sat}(\cdot)$ was defined as follows:
\begin{equation}
    \bar{u}_{k} = f_\text{sat}(u_{k}) = 0.5 u_{k}\left(1+\text{sign}(u_{k})\right), 
\end{equation}
where $\text{sign}(\cdot)$ denotes the sign function.

We determined the parameter settings by trial and error. The reservoir size $r$ was set to 50, and the leaky rate $\gamma$ was 0.8. The reservoir $\mathbf{W}^\text{r}$ was initialized with $\mathcal{U}(-0.5, 0.5)$ elements, where $\mathcal{U}(a, b)$ denotes a random value following a uniform distribution between $a$ and $b$, $a<b$. After initialization, $\mathbf{W}^\text{r}$ was scaled such that its largest absolute eigenvalue, which is known as the spectral radius, was 0.8. The input layer $\mathbf{W}^\text{in}$ was randomly initialized by $\mathcal{U}(-1, 1)$ elements and then scaled by a factor of 0.1. We set the tap size $\delta = 5$ and initialized the states of both ESN-C and ESN-L (i.e., $\mathbf{x}^\text{C}$ and $\mathbf{x}^\text{L}$) by running (\ref{eq: ESN-C update}) and (\ref{eq: ESN-L update}) with zero input for 100 steps. The output layer $\mathbf{w}^\text{out}$ was initialized with $\mathcal{N}(0, 1)$ elements and scaled by 0.01. The learning rate $\alpha$ was 1, and the forgetting factor $\lambda$ was $1-1\times10^{-3}$. The P-D controller's P gain $K_P$ was  $1\times10^{-3}$, and the D gain $K_D$ was $1\times10^{-5}$.

For performance comparison, we evaluated three additional control methods: the online learning ESN control system without the P-D controller (denoted as ESN), the pre-trained online learning ESN control system without the P-D controller (denoted as TESN), and the P-D controller without the online learning ESN (denoted as FB). The TESN method's prior training was conducted following \cite{park2016online} by adjusting the output layer $\mathbf{w}^\text{out}$ with the RLS algorithm on a dataset obtained by feeding 100 steps of $f_\text{sat}\left(\mathcal{N}(1, 0.3)\right)$ input into the system. The control system shown in Fig.~\ref{fig: system}, which combines the online learning ESN controller with the P-D controller, is referred to as ESN+FB in the subsequent discussion. For both the ESN and the TESN methods, due to the lack of $u^\text{b}$, the control rule in (\ref{eq: ESN+FB output}) is simplified to:
\begin{equation}
    u_{k} = u^\text{f}_{k}.
\label{eq: TESN output}
\end{equation}
Similarly, for the FB model, the control output is given by:
\begin{equation}
    u_{k} = \bar{u}_{k-1} + u^\text{b}_{k}.
\label{eq: FB output}
\end{equation}
Each simulation in this work was repeated with 1,000 pre-determined and different random seeds. We use solid lines to illustrate the average across 1,000 simulation results and shaded regions to show the average $\pm$ the standard deviation.

\subsection{Performance Comparison}

\begin{figure}[tb]
    \centering
\includegraphics[width=\linewidth]{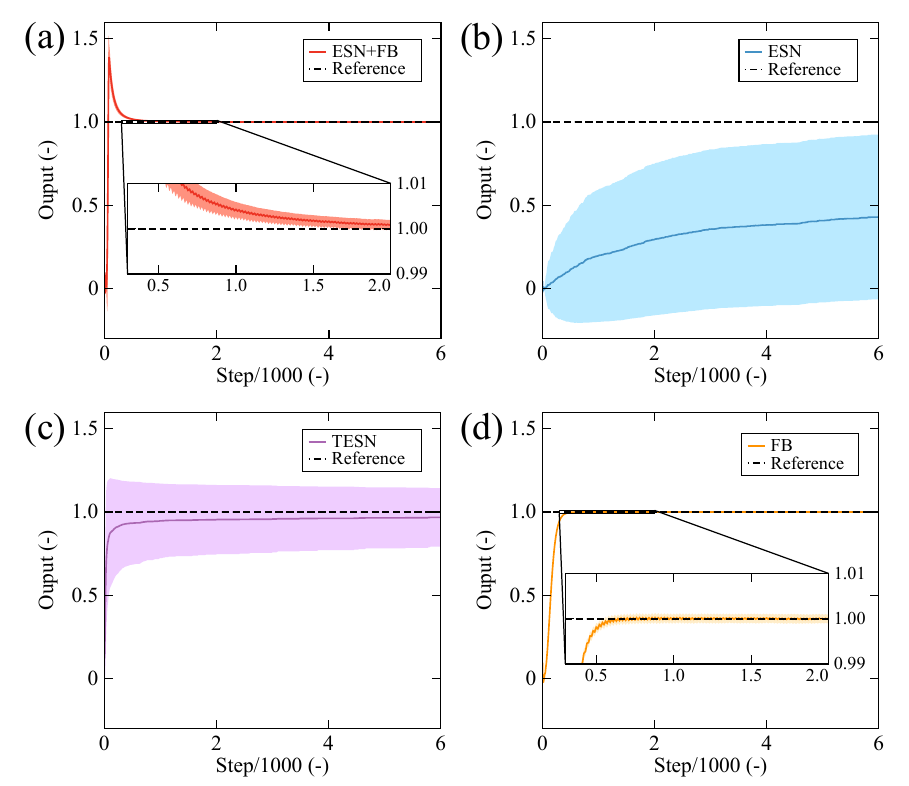}
    \caption{Step responses: (a) ESN+FB; (b) ESN; (c) TESN; (d) FB.}
    \label{fig: step}
\end{figure}

We compared the performance of different control methods in tracking a step response signal, with results shown in Fig.~\ref{fig: step}. The results show that both the ESN+FB and the FB methods successfully followed the step signal. The FB exhibited better transient behavior than the ESN+FB, with faster convergence and no overshooting. The ESN method failed to track this step signal due to its slow convergence. The TESN method also showed unsatisfactory performance, with control errors not fully converging to 0 and the broad uncertainty regions indicated by the shadow areas.

\begin{figure}[tb]
    \centering
\includegraphics[width=\linewidth]{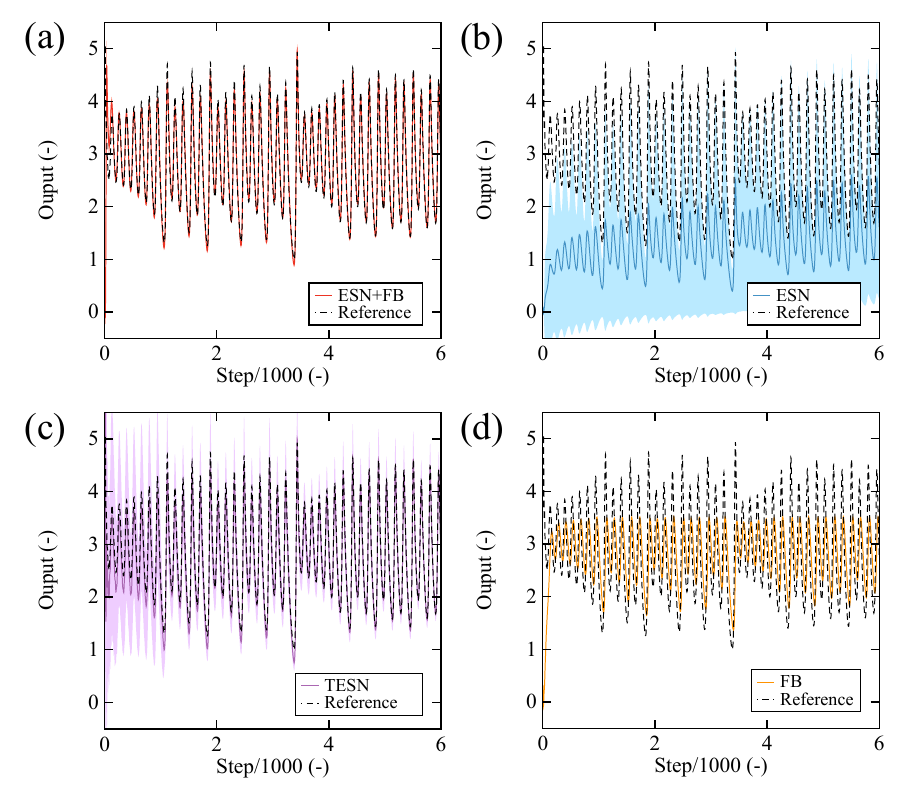}
    \caption{Tracking results: (a) ESN+FB; (b) ESN; (c) TESN; (d) FB.}
    \label{fig: tracking}
\end{figure}

We tested different control methods in tracking a complex varying signal, with results shown in Fig.~\ref{fig: tracking}. The results show that the ESN+FB method closely followed the reference signal, whereas the FB method failed due to the absence of the feedforward element $u^\text{f}$. The ESN method also failed to track this reference signal because of its slow convergence. Though the TESN method succeeded in this control task, it exhibited poorer transient performance than the ESN+FB, with notably broader uncertainty regions at the initial stage.

\begin{figure}[tb]
    \centering
\includegraphics[width=\linewidth]{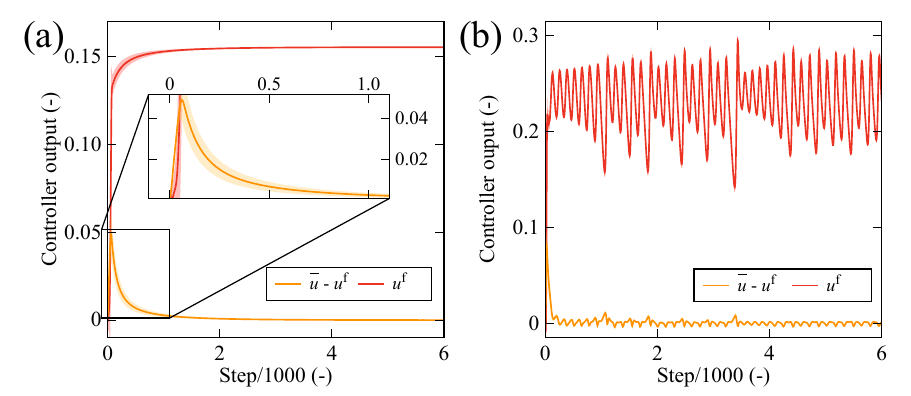}
    \caption{Feedforward and feedback control elements of the ESN+FB method: (a) Tracking the step response signal; (b) Tracking the complex signal.}
    \label{fig: output}
\end{figure}

The feedforward control element $u^\text{f}$ and the feedback element, defined as $\bar{u} - u^\text{f}$, of the ESN+FB control system during the above two control tasks are shown in Fig.~\ref{fig: output}. The results indicate that while introducing the feedback controller realizes the ESN+FB method's greatly enhanced performance compared to the ESN method, the feedback element plays a relatively important role in the total control output $\bar{u}$ merely during the initial transient phase. After which, the feedback element contributes only a negligible portion to $\bar{u}$, consistent with the findings in \cite{xing2012modeling}. Hence, the P-D controller's role in the ESN+FB system is considered primarily to accelerate the convergence of the online learning ESN controller rather than to directly control the plant.

\begin{figure}[tb]
    \centering
\includegraphics[width=\linewidth]{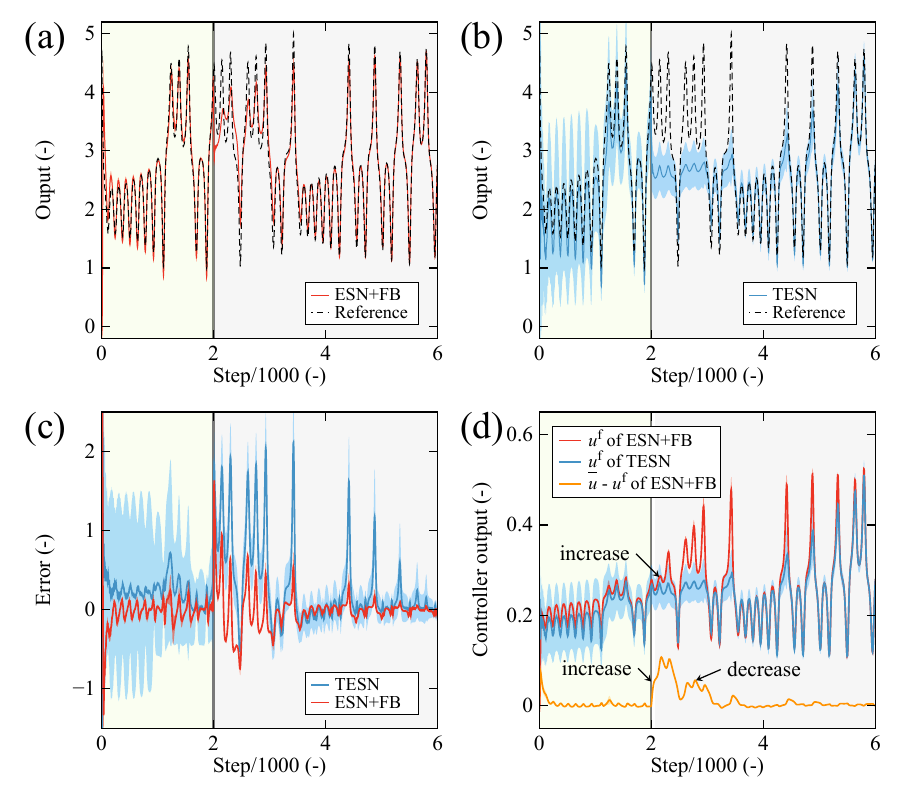}
    \caption{Tracking a complex signal with system changes: (a) ESN+FB; (b) TESN; (c) Control errors; (d) Controller outputs. Green zones before the 2,000th step denotes the system was unchanged as (\ref{eq: Nonlinear 1}), grey zones after the 2,000th step means the system's dynamics has changed to (\ref{eq: Nonlinear 2}).}
    \label{fig: disturbance}
\end{figure}

We compared the robustness of the ESN+FB and TESN control methods against external disturbances while tracking a complex reference signal. The external disturbance was simulated by changing the dynamics of the controlled system. Specifically, before the 2,000th step, the system's dynamics followed (\ref{eq: Nonlinear 1}). Afterward, the dynamics was changed to:
\begin{equation}
    {y_{k+1}}^3 = \frac{{y_{k}}^8}{7+{y_{k-1}}^6} + \left( 9 \tanh(\bar{u}_k) \right)^3.
\label{eq: Nonlinear 2}
\end{equation}
The test results are shown in Fig.~\ref{fig: disturbance}. Before the system change, the ESN+FB exhibited fewer control errors and negligible shadow regions compared to the TESN. Both methods experienced significant control errors following the sudden change in plant's dynamics. As highlighted by the arrows in Fig.~\ref{fig: disturbance}(d), after the change, the ESN+FB system's feedback element showed a sharp increase, which prompted the RLS algorithm's adaptation to the new dynamics, thus leading to a corresponding increase in the feedforward element, after which the feedback element gradually returned to a steady low level. In contrast, the TESN's output exhibited a slower adaption to the system change, requiring roughly 4,000 steps to recover and stabilize the errors, while the ESN+FB system recovered in around 2,000 steps, as shown by Fig.~\ref{fig: disturbance}(c).

\begin{figure}[tb]
    \centering
\includegraphics[width=\linewidth]{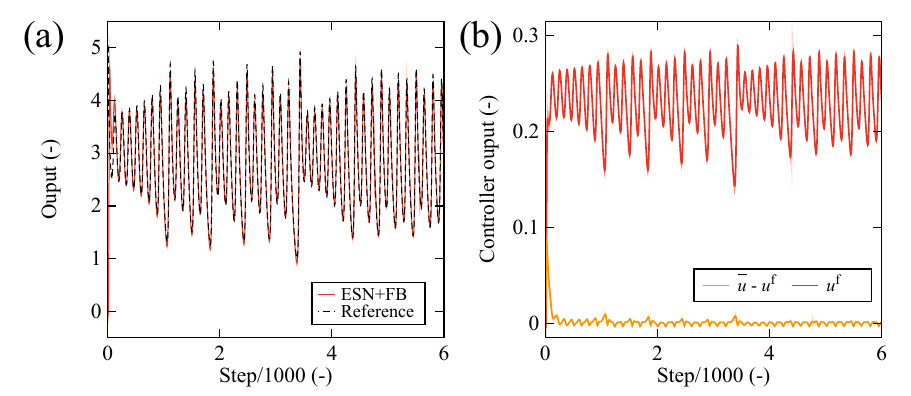}
    \caption{Tracking a complex signal using the ESN+FB method with random RLS hyperparameter settings: (a) Tracking result; (b) Controller outputs.}
    \label{fig: random_rls}
\end{figure}

\begin{figure}[tb]
    \centering
\includegraphics[width=\linewidth]{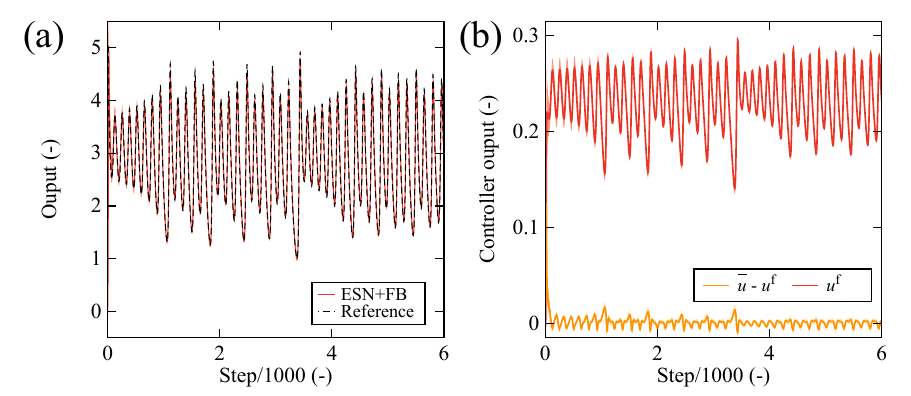}
    \caption{Tracking a complex signal using the ESN+FB method with random feedback P-D controller gains: (a) Tracking result; (b) Controller outputs.}
    \label{fig: random_pd}
\end{figure}

The performance of machine learning models can be sensitive to hyperparameter settings. We evaluated the ESN+FB method's robustness to variations in hyperparameter setups in controlling the nonlinear system described by (\ref{eq: Nonlinear 1}) to track a varying signal. First, in each simulation run, the RLS learning rate $\alpha$ and forgetting factor $\lambda$ were randomized as follows:
\begin{align}
    \alpha &= 10^{\mathcal{U}(-1, 1)}  \label{eq: random alpha} \\
    \lambda &= 1-10^{\mathcal{U}(-4, -2)}  \label{eq: random lambda}
\end{align}
with all other parameters fixed as per the original setup. The results are presented in Fig.~\ref{fig: random_rls}. Next, we tested the ESN+FB method with the proportional and differential gains set as:
\begin{align}
    K_P &= \mathcal{U}(0, 0.01) \\
    K_D &= \mathcal{U}(0, 1\times10^{-4})
\end{align}
with all other parameters unchanged. The results are shown in Fig.~\ref{fig: random_pd}. As Fig.~\ref{fig: random_rls} and Fig.~\ref{fig: random_pd} show, the ESN+FB method maintained a similar control performance as Fig.~\ref{fig: tracking}, despite random variations in the RLS setups or feedback gains, implying that it can be easily implemented with minimal hyperparameter tuning. Additionally, we observed that the ESN output $u^\text{f}$ in Fig.~\ref{fig: random_pd}(b) exhibited slightly wider uncertainty regions during the initial phase (approximately the first 1,000 steps) compared to Fig.~\ref{fig: output}(b), meaning that variations in feedback gains primarily affect the convergence behavior of the ESN+FB control system at the initial transient stage.

\begin{figure}[tb]
    \centering
\includegraphics[width=\linewidth]{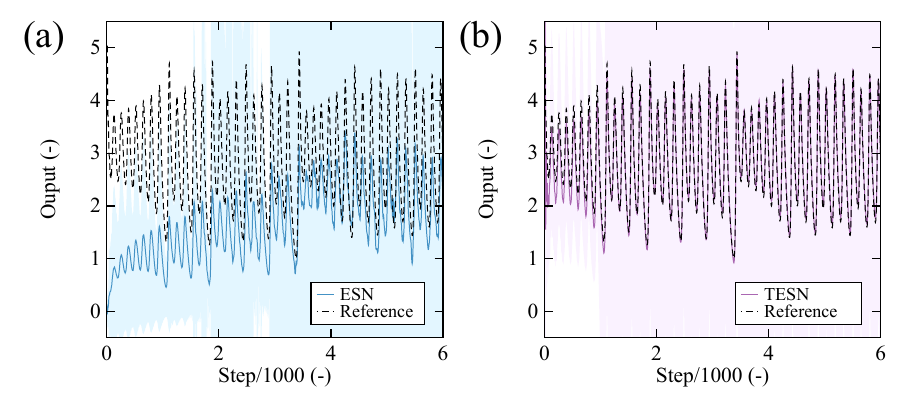}
    \caption{Tracking a complex signal using the ESN and the TESN methods with random RLS hyperparameter settings: (a) ESN result; (b) TESN result.}
    \label{fig: random_rls_esn}
\end{figure}

For comparison, we tested the ESN and TESN methods with the same random RLS setups as defined in (\ref{eq: random alpha}) and (\ref{eq: random lambda}). The results, presented in Fig.~\ref{fig: random_rls_esn}, show that both the ESN and TESN methods are sensitive to hyperparameter variations, exhibiting unstable performance when the RLS hyperparameters were altered. This instability is characterized by the greatly broader uncertainty regions than those in Fig.~\ref{fig: tracking}.

\section{Discussion}

The training data distribution can significantly impact the machined learning model's performance in specific tasks \cite{hammoudeh2024training}. A model tends to perform better when there is alignment between the distribution of its training data and the data distribution encountered in practice, and vice versa \cite{hammoudeh2024training}. Here, we define the optimal input-output data distribution of the online learning ESN control system given a specific reference signal as the distribution of the controlled system's input-output data when its output matches or closely follows the reference signal. In this section, we use simulations to demonstrate that the ESN+FB method's accelerated convergence stems from the feedback P-D controller guiding the RLS algorithm to adjust the ESN model with input-output training data closer to the optimal data distribution. In the following content, the ESN+FB control system's output $\bar{u}$ is used as the optimal input data to the specific reference signal.

\begin{figure}[tb]
    \centering
\includegraphics[width=\linewidth]{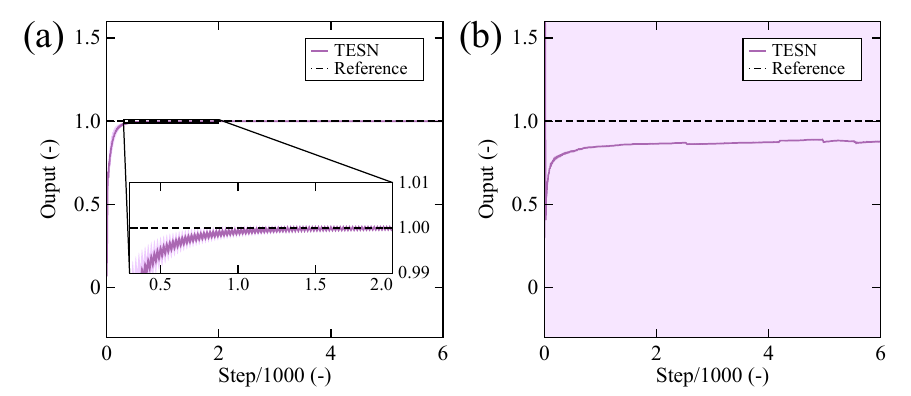}
    \caption{Tracking the step response signal with TESN methods: (a) The ESN model underwent the prior training with $f_\text{sat}\left(\mathcal{N}(0.15, 0.05)\right)$; (b) The ESN model underwent the prior training with $f_\text{sat}\left(\mathcal{N}(2.4, 0.8)\right)$.}
    \label{fig: step_distribution}
\end{figure}

We used the TESN method to control the system described by (\ref{eq: Nonlinear 1}) following a step response signal. Specifically, we modified the TESN model's original training data $f_\text{sat}\left(\mathcal{N}(1.0, 0.3)\right)$ to $f_\text{sat}\left(\mathcal{N}(0.15, 0.05)\right)$, which is closer to the distribution of the optimal input shown in Fig.~\ref{fig: output}(a), and also to $f_\text{sat}\left(\mathcal{N}(2.4, 0.8)\right)$, which deviates further from the optimal distribution than the original training data. The 100-step training amount remained consistent. The results are shown in Fig.~\ref{fig: step_distribution}. The comparison between Fig.~\ref{fig: step_distribution}(a) and Fig.~\ref{fig: step}(c) shows that the TESN performs better when trained with data closer to the optimal distribution, and vice versa, as shown by the comparison between Fig.~\ref{fig: step}(c) and Fig.~\ref{fig: step_distribution}(b).

\begin{figure}[tb]
    \centering
\includegraphics[width=\linewidth]{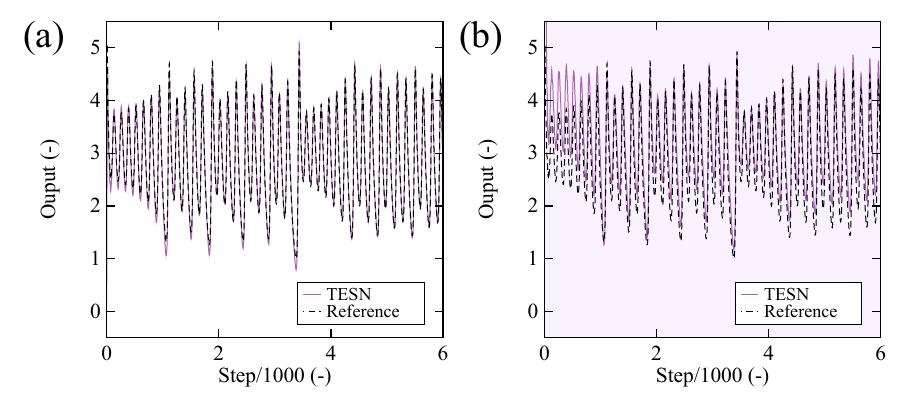}
    \caption{Tracking the complex reference signal with TESN methods: (a) The ESN model underwent the prior training with $f_\text{sat}\left(\mathcal{N}(0.24, 0.08)\right)$; (b) The ESN model underwent the prior training with $f_\text{sat}\left(\mathcal{N}(1.5, 0.5)\right)$.}
    \label{fig: tracking_distribution}
\end{figure}

Additionally, we conducted the test in Fig.~\ref{fig: tracking} by the TESN trained on the data $f_\text{sat}\left(\mathcal{N}(0.24, 0.08)\right)$, which is closer to the distribution shown in Fig.~\ref{fig: output}(b), and $f_\text{sat}\left(\mathcal{N}(1.5, 0.5)\right)$, which deviates further from the distribution in Fig.~\ref{fig: output}(b) compared to the original training data $f_\text{sat}\left(\mathcal{N}(1.0, 0.3)\right)$. The results are presented in Fig.~\ref{fig: tracking_distribution}. The control outcome in Fig.~\ref{fig: tracking_distribution}(a) shows faster convergence and smaller uncertainty regions compared to Fig.~\ref{fig: tracking}(c). In contrast, the result in Fig.~\ref{fig: tracking_distribution}(b) demonstrates worse performance featured by significantly broader shadow regions than Fig.~\ref{fig: tracking}(c).

The above analyses demonstrate that when the training data is closer to the optimal distribution, the TESN performs better, thus underscoring the importance of the training data distribution for the performance of the TESN online learning control system. Although the sole P-D controller could not track the varying signal, as shown in Fig.~\ref{fig: tracking}(d), its sensitivity and swift response to control errors effectively guided the controlled system's input-output closer to the optimal data distribution. This alignment allowed the RLS algorithm to adjust the ESN model using training data highly similar to the optimal data corresponding to the given control task (i.e., the ground truth test data), thus accelerating the online learning ESN control system's convergence in specific missions.

\begin{figure}[tb]
\centering
\includegraphics[width=\linewidth]{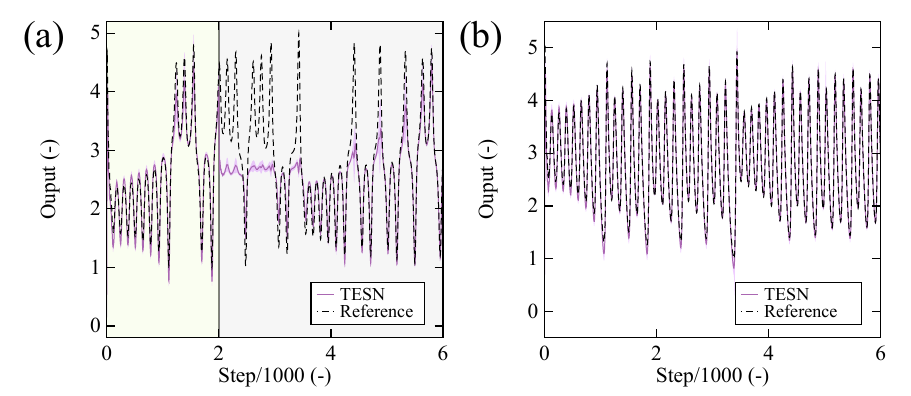}
\caption{Robustness tests of the TESN with the ESN trained on the dataset $f_\text{sat}\left(\mathcal{N}(0.24, 0.08)\right)$: (a) The test in Fig.\ref{fig: disturbance}; (b) The test in Fig.\ref{fig: random_rls_esn}.}
\label{fig: disturbance_distribution}
\end{figure}

We conducted the robustness tests shown in Fig.~\ref{fig: disturbance} and Fig.~\ref{fig: random_rls_esn} using the TESN trained with $f_\text{sat}\left(\mathcal{N}(0.24, 0.08)\right)$. The results are shown in Fig.~\ref{fig: disturbance_distribution}(a) and (b), respectively. As observed in Fig.~\ref{fig: disturbance_distribution}(a), with a training set's distribution closer to the optimal dataset, the TESN closely followed the reference signal before the system's change. However, the TESN still showed a slow recovery from the sudden change in system dynamics. In contrast, as shown in Fig.~\ref{fig: disturbance}, the feedback controller in the ESN+FB system responded rapidly to the system's change, driving the online learning ESN into the changed optimal zone corresponding to the altered system dynamics, thus enabling the ESN+FB system's adaption to the sudden change. The comparison between Fig.~\ref{fig: disturbance_distribution}(b) and Fig.~\ref{fig: random_rls_esn}(b) indicates that with better-selected training data, the TESN system demonstrated stronger robustness against variations in the RLS algorithm's hyperparameter settings.

Selecting a suitable dataset for the TESN's prior training needs expertise and can be particularly challenging without prior knowledge of the system's dynamics and the reference signal's features. Furthermore, even with a well-designed dataset for prior training, the TESN still performs poorly in the presence of external disturbances, such as changes in the controlled system's dynamics, as shown in Fig.~\ref{fig: disturbance_distribution}(a). In contrast, the ESN+FB method can be employed in a zero-shot manner, free from these limitations. The feedback controller can consistently guide the online learning ESN to the optimal training data field, facilitating fast convergence regardless of the reference signals or plant dynamics. Moreover, despite the substantial performance improvements introduced by the feedback controller, its structure and mechanism are much simpler than that of the online learning ESN controller, making it easy to integrate into any existing online learning ESN control system with minimal effort.

We also observed that while the feedback controller accelerates the online learning ESN's convergence, the completely untrained ESN model can produce undesired outputs, such as significant overshooting during the initial phase, as seen in Fig.~\ref{fig: step}(a) and Fig.~\ref{fig: tracking}(a). Strategies such as constraining the randomly initialized and untrained ESN model's output at a very early period can be considered to prevent this undesired large overshooting in practical applications.

\section{Conclusion}

This work demonstrated that incorporating a feedback controller can effectively address the slow convergence issue in RLS-based online learning ESN control systems. Simulations verified the feedback controller’s ability to accelerate the convergence of the online learning ESN system. The proposed control method also exhibited strong robustness against variations in both the controlled plant’s dynamics and the online learning controller’s hyperparameters. We showed that the feedback controller enhances the convergence of the online learning ESN by guiding the RLS algorithm to adjust the ESN model within a data field closer to the distribution of the ground truth test data. This study is expected to offer practical benefits for engineers implementing online learning ESN control systems.

\FloatBarrier


\begin{thebibliography}{10}
\providecommand{\url}[1]{#1}
\csname url@rmstyle\endcsname
\providecommand{\newblock}{\relax}
\providecommand{\bibinfo}[2]{#2}
\providecommand\BIBentrySTDinterwordspacing{\spaceskip=0pt\relax}
\providecommand\BIBentryALTinterwordstretchfactor{4}
\providecommand\BIBentryALTinterwordspacing{\spaceskip=\fontdimen2\font plus
\BIBentryALTinterwordstretchfactor\fontdimen3\font minus \fontdimen4\font\relax}
\providecommand\BIBforeignlanguage[2]{{%
\expandafter\ifx\csname l@#1\endcsname\relax
\typeout{** WARNING: IEEEtran.bst: No hyphenation pattern has been}%
\typeout{** loaded for the language `#1'. Using the pattern for}%
\typeout{** the default language instead.}%
\else
\language=\csname l@#1\endcsname
\fi
#2}}

\bibitem{cheng2016adaptive}
L.~Cheng, W.~Liu, Z.-G. Hou, T.~Huang, J.~Yu, and M.~Tan, ``An adaptive takagi--sugeno fuzzy model-based predictive controller for piezoelectric actuators,'' \emph{IEEE Trans. Ind. Electron.}, vol.~64, no.~4, pp. 3048--3058, 2016.

\bibitem{zou2023generalized}
J.~Zou, S.~O. Kassim, J.~Ren, V.~Vaziri, S.~S. Aphale, and G.~Gu, ``A generalized motion control framework of dielectric elastomer actuators: dynamic modeling, sliding-mode control and experimental evaluation,'' \emph{IEEE Trans. Robot.}, 2023.

\bibitem{waegeman2012feedback}
T.~Waegeman, B.~Schrauwen, \emph{et~al.}, ``Feedback control by online learning an inverse model,'' \emph{IEEE Trans. Neural Netw. Learn. Syst.}, vol.~23, no.~10, pp. 1637--1648, 2012.

\bibitem{liang2023online}
J.~Liang, Z.~Ding, Q.~Han, H.~Wu, and J.~Ji, ``Online learning compensation control of an electro-hydraulic shaking table using echo state networks,'' \emph{Eng. Appl. Artif. Intell.}, vol. 123, p. 106274, 2023.

\bibitem{sun2022physics}
W.~Sun, N.~Akashi, Y.~Kuniyoshi, and K.~Nakajima, ``Physics-informed recurrent neural networks for soft pneumatic actuators,'' \emph{IEEE Robot. Autom. Lett.}, vol.~7, no.~3, pp. 6862--6869, 2022.

\bibitem{jiang2022intelligent}
Z.~Jiang, Y.~Li, and Q.~Wang, ``Intelligent feedforward hysteresis compensation and tracking control of dielectric electro-active polymer actuator,'' \emph{Sens. Actuators A: Phys.}, vol. 341, p. 113581, 2022.

\bibitem{jaeger2002adaptive}
H.~Jaeger, ``Adaptive nonlinear system identification with echo state networks,'' \emph{Adv. Neural Inf. Process. Syst.}, vol.~15, 2002.

\bibitem{park2016online}
J.~Park, B.~Lee, S.~Kang, P.~Y. Kim, and H.~J. Kim, ``Online learning control of hydraulic excavators based on echo-state networks,'' \emph{IEEE Trans. Autom. Sci. Eng.}, vol.~14, no.~1, pp. 249--259, 2016.

\bibitem{xing2012modeling}
K.~Xing, Y.~Wang, Q.~Zhu, and H.~Zhou, ``Modeling and control of mckibben artificial muscle enhanced with echo state networks,'' \emph{Control Eng. Pract.}, vol.~20, no.~5, pp. 477--488, 2012.

\bibitem{jordanou2019online}
J.~P. Jordanou, E.~A. Antonelo, and E.~Camponogara, ``Online learning control with echo state networks of an oil production platform,'' \emph{Eng. Appl. Artif. Intell.}, vol.~85, pp. 214--228, 2019.

\bibitem{zhou2020deep}
S.~Zhou, M.~K. Helwa, and A.~P. Schoellig, ``Deep neural networks as add-on modules for enhancing robot performance in impromptu trajectory tracking,'' \emph{Int. J. Robot. Res.}, vol.~39, no.~12, pp. 1397--1418, 2020.

\bibitem{hammoudeh2024training}
Z.~Hammoudeh and D.~Lowd, ``Training data influence analysis and estimation: A survey,'' \emph{Mach. Learn.}, vol. 113, no.~5, pp. 2351--2403, 2024.

\end{thebibliography}
\end{document}